\documentclass[11pt,showpacs]{revtex4}
\usepackage{graphicx}
\usepackage{epsfig}
\usepackage{amsmath,amssymb}
\def\eqref#1{Eq.~(\ref{eq:#1})}

\def\figref#1{Fig.~(\ref{fig:#1})}

\begin{document}

\title{
Dilepton production at HADES: theoretical predictions
}
\author{M.D. Cozma}
\affiliation{Institut f\"{u}r Theoretische Physik, 
Universit\"{a}t T\"{u}bingen, Auf der Morgenstelle 14, 72076
  T\"{u}bingen, Germany}
\author{C. Fuchs}
\affiliation{Institut f\"{u}r Theoretische Physik, 
Universit\"{a}t T\"{u}bingen, Auf der Morgenstelle 14, 72076
  T\"{u}bingen, Germany}
\author{E. Santini}
\affiliation{Institut f\"{u}r Theoretische Physik, 
Universit\"{a}t T\"{u}bingen, Auf der Morgenstelle 14, 72076
  T\"{u}bingen, Germany}
\author{A. F\"{a}ssler}
\affiliation{Institut f\"{u}r Theoretische Physik, 
Universit\"{a}t T\"{u}bingen, Auf der Morgenstelle 14, 72076
  T\"{u}bingen, Germany}

\begin{abstract}
Dileptons represent a unique probe for nuclear matter under extreme
conditions reached in heavy-ion collisions. They allow to study
meson properties, like mass and decay width, at various density and temperature
regimes. Present days models allow generally a good description
of dilepton spectra in ultra-relativistic heavy ion collision. For the energy regime
of a few GeV/nucleon, important discrepancies between theory and experiment, known as
the DLS puzzle, have been observed. Various models, including the one developed by the
T\"{u}bingen group, have tried to address this problem, but have proven only partially
successful. High precision spectra of dilepton emission in heavy-ion reactions at 1 and
2 GeV/nucleon will be released in the near future by the HADES Collaboration at GSI. 
Here we present the predictions for dilepton spectra in C+C reactions 
at 1 and 2 GeV/nucleon and investigate up to what degree possible scenarios for the in-medium
modification of vector mesons properties are accessible by the HADES experiment. 

\end{abstract}
\maketitle

\section{Introduction}

Heavy ion reactions present an unique opportunity for the study of nuclear matter under extreme
conditions allowing a comprehensive analysis of the phase structure of the underlying theory
of strong interactions. In this process electromagnetic probes such as dileptons have 
been proven to be most effective since they leave the reaction zone 
essentially undistorted by final state interactions. They provide thus  a clear
view on effective degrees of freedom at high baryon density and temperature. It has been argued
that their differential spectra could reveal information about chiral restoration and in-medium
properties of hadrons~\cite{Bro91,Hat92,Sha94}. 
Theoretically, there exists an abundance of models that 
predict a change of vector meson masses and
widths in high density/temperature nuclear matter: 
Brown-Rho scaling~\cite{Bro91} is equivalent
with a decrease of vector meson masses in nuclear medium; models based on QCD sum rules~\cite{Hat92}
and effective hadronic models~\cite{Her93,Fri97} reach similar conclusions.
\par
Experimentally, dilepton spectra have been measured at two different energy regimes: 
the CERES~\cite{Aga95,ceres2}, HELIOS~\cite{Maz94} and recently NA60 \cite{na60} 
at CERN have measured dielectrons and dimuons, 
respectively, in heavy ion reactions at 158 GeV/nucleon. 
 In proton-nucleus reactions 
the sum over all measured hadronic sources, 
i.e. the so-called hadronic cocktail, describes the corresponding 
dilepton spectra perfectly well. However,  in heavy systems (Pb+Au)
a significant enhancement of the dilepton spectra below the the $\rho$ and $\omega$ peaks
has been observed relative to the corresponding hadronic 
cocktail. Such a behaviour could
be explained theoretically, within a scenario of a 
dropping $\rho$ vector meson mass~\cite{Cas95} or by the
inclusion of in-medium spectral functions for the 
vector mesons~\cite{Urb98,Bra98}. The recent NA60 dimuon 
spectra with high resolution in the vicinity around the 
$\rho-\omega$ peak seem to rule out a naive dropping mass 
scenario but support the picture of modified $\rho-\omega$ spectral 
functions. An enhanced strength below the $\omega$ peak has 
also been observed in $\gamma$-nucleus reactions \cite{cbelsa}.  
A second set of heavy ion experiments
have been performed at laboratory energies of 1.0 AGeV (Ca+Ca and C+C) by the DLS
Collaboration at BEVALAC~\cite{Por97,DLS2}. Also in this case, the low mass 
region of the dilepton spectra is underestimated by present transport 
calculations, in contrast with similar measurements (1.04 - 4.88 GeV/nucleon)
for the p+p and p+d systems. As opposed to the ultra-relativistic case, 
the situation does not improve when the
in-medium spectral functions or the dropping mass scenarios 
are taken into account~\cite{Bra98,Ern98} (the DSL
puzzle). Other scenarios like possible contributions from the 
quark-gluon plasma or in-medium modifications
of the $\eta$ mass have been excluded as a possible resolution 
of this puzzle. Decoherence effects~\cite{She03}
have been proven to be partially successful in explaining the difference 
between the DLS data and the theoretical predictions.
\par
Recently, a new measurement by the HADES Collaboration at GSI has been completed and
the results will be published in the near future~\cite{Had05}. 
The aim of this second generation experiment
is to measure dilepton spectra in A+A, p+A and $\pi$+A reactions 
with an unprecedented mass resolution ($\Delta M/M\simeq 1\%(\sigma)$) over 
the entire spectrum \cite{Fri99}. Such a resolution allows to 
measure the in-medium properties (mass and width) 
of $\rho$ and $\omega$ mesons through their decays into 
dielectron pairs in nuclear matter with high precision and will put 
strong constraints on   theoretical models. 
This  letter presents predictions of the dilepton production in C+C reactions 
at 1.0 and 2.0 AGeV which are those reactions where first data from HADES will 
be available in the near future. The vector meson and dilepton production 
is described within the framework of the resonance model developed in 
\cite{Fae00,Kri01,Fae03} in combination 
with the relativistic quantum molecular dynamics (RQMD) transport 
model for  heavy ion collisions \cite{She03}. The influence of medium 
effects such as quantum decoherence, collisional broadening and a 
dropping vector meson mass are investigated. 
The paper is organised as follows: in Section II we give a brief 
description of the elementary reactions
which contribute to dilepton emission in heavy ion collisions, 
together with an outline of the RQMD model.
Section III is devoted to the presentation of our prediction for 
the differential mass spectrum  of dilepton
production in 1.0 and 2.0 AGeV C+C collisions. We conclude 
with Section IV.

\section{The model}

\subsection{Elementary dilepton sources}

\begin{figure*}
\centering
\includegraphics[width=11cm]{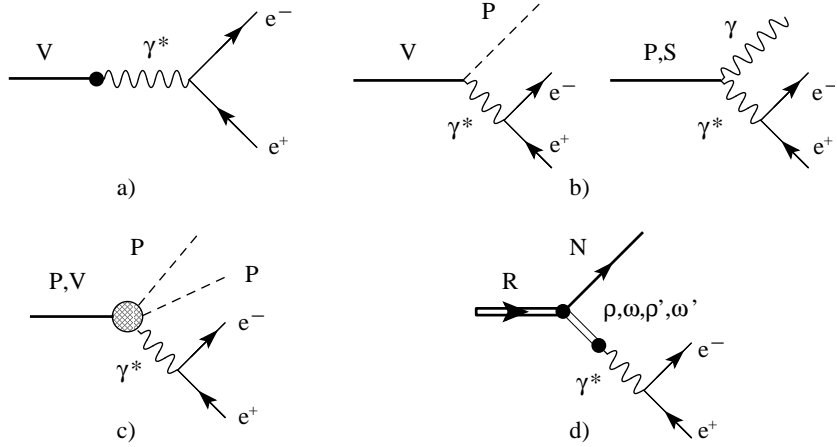}
\caption{\label{fig:fey_diags} Feynman diagrams of the elementary processes contributing to
dilepton emission: a) direct decay of a vector meson ($\rho,\omega,\phi$) to a dilepton pair
going through an intermediate photon (VMD model); b) Dalitz decays of a vector (V), pseudo-scalar (P)
or scalar meson (S) into a dilepton pair and a photon or a pseudo-scalar meson ($\eta$ or $\pi$);c) four-body
decay into a dilepton pair plus two pseudo-scalar mesons (the hashed vertex represents an intermediate state
containing a vector mesons and/or virtual photon - see Ref.~\cite{Fae00}); and d) the decay of a nucleon or
$\Delta$ resonance into a nucleon plus a virtual vector meson (extended VMD) which then decays into two dileptons.}
\end{figure*}

The elementary sources of dilepton production in heavy ion reaction in the energy range of a few GeV/nucleon
are numerous. One can identify three main classes of processes that lead to dilepton emission: nucleon-nucleon bremsstrahlung, 
decays of light unflavoured mesons and decay of nucleon and $\Delta$ resonances. For the energy
range of interest in this paper dilepton generation through nucleon-nucleon bremsstrahlung is unimportant.
Feynman diagrams of processes belonging to the last two classes are depicted in ~\figref{fey_diags}.

At incident energies of 1 AGeV the cross-sections for meson production $\mathcal{M}=\eta,\eta',
\rho,\omega,\phi$ are small and these mesons do not play an important role in the dynamics of heavy-ion
collisions. Their production can thus be treated perturbatively, in contrast to the case of the pion. The
decay to a dilepton pair takes place through the emission of a virtual photon. The differential branching
ratios for the decay of a meson to a final state $Xe^+e^-$ can be written
\begin{eqnarray}
dB(\mu,M)^{\mathcal{M},\pi\rightarrow e^+e^-X}=\frac{d\Gamma(\mu,M)^{\mathcal{M},\pi\rightarrow e^+e^-X}}
{\Gamma_{tot}^{\mathcal{M},\pi}(\mu)}\:,
\end{eqnarray}
with $\mu$ the meson mass and $M$ the dilepton mass. Three types of such decays have been considered: direct
decays $\mathcal{M}\rightarrow e^{+}e^{-}$ (Fig.~\ref{fig:fey_diags}a), Dalitz decays $\mathcal{M}\rightarrow
\gamma e^{+}e^{-}$, $\mathcal{M}\rightarrow\pi(\eta) e^{+}e^{-}$ (Fig.~\ref{fig:fey_diags}b) and four-body decays
$\mathcal{M}\rightarrow\pi\pi e^{+}e^{-}$ (Fig.~\ref{fig:fey_diags}c). A comprehensive study of the decay of
light mesons to a dilepton pair has been performed in~\cite{Fae00}, the decay channels there are most important
quantitatively for heavy-ion collisions at 1 and 2 GeV/nucleon being $\pi^0\rightarrow \gamma e^+e^-$ and
$\eta\rightarrow \gamma e^+e^-$.

The third source for dilepton emission we have mentioned 
was the decay of baryonic resonances (see Fig.~\ref{fig:fey_diags}d). 
For the description of this process an extension of the vector meson dominance (VMD)
model has been employed~\cite{Fae00,Kri01}. The original VMD model 
assumes that decays of baryon resonances run through an intermediate 
virtual meson ($\rho$ or $\omega$) required for the description of the 
form-factors entering in the calculation of the radiative 
($R\rightarrow N\gamma$) and mesonic ($R\rightarrow NV$) decays. 
Such a model does not allow the simultaneous description of both 
radiative and mesonic decays~\cite{Fri97,Fae03,Pos01}. Furthermore
the quark counting rules require a stronger suppression of the 
transition form-factor than the $1/t$ behaviour predicted
by the naive VMD. Similarly the $\omega\pi\gamma$ transition 
form-factor shows an asymptotic $1/t^2$ behaviour~\cite{Vai78}. 
An extension of the VMD to allow contributions from radially 
excited vector mesons ($\rho$(1250), $\rho$(1450),$\ldots$ in 
Ref.~\cite{Kri01}) that interfere destructively with the ground state vector mesons
($\rho$ in this example) allow for a resolution of the mentioned problems of the original 
VMD and describe the radiative and mesonic decays in a unitary way.

In terms of the branching ratios for the Dalitz decays of the baryon resonances, the cross section for $e^+ e^-$
production from the initial state $X'$ together with the final state $NX$ can be written
\begin{eqnarray}
\frac{d\sigma(s,M)^{X'\rightarrow NXe^+e^-}}{dM^2}=\sum_R\int_{(m_N+M)^2}^{\sqrt{s}-m_X)^2}\:
d\mu^2\\
\times\,\frac{d\sigma(s,\mu)^{X'\rightarrow RX}}{d\mu^2}\:
\sum_V\,\frac{dB(\mu,M)^{R\rightarrow VN \rightarrow Ne^+e^-}}{dM^2}\,, \nonumber
\end{eqnarray}
where $\mu$ is the mass of the baryon resonance $R$ which has the production cross-section
$d\sigma(s,\mu)^{X'\rightarrow XR}$ and $dB(\mu,M)^{R\rightarrow VN \rightarrow e^+e^-}$ being
the differential branching ratio for the decay of the resonance $R\rightarrow Ne^+e^-$ through the
vector meson $V$. The initial state $X'$ could consist of two baryons ($X'=NN,NR,RR'$) or of one
nucleon and a pion ($X=\pi N$). The dilepton decay rate can be found, once the width $\Gamma(R\rightarrow
N\gamma^*$) is known  by using the factorisation prescription
\begin{eqnarray}
d\Gamma(R\rightarrow Ne^+e^-)&=&\Gamma(R\rightarrow N\gamma^*)\,M\,\Gamma(\gamma^*\rightarrow e^+e^-)
\frac{dM^2}{\pi M^4}\,,
\end{eqnarray}
with
\begin{eqnarray}
M\Gamma(\gamma^*\rightarrow e^+e^-)&=&\frac{\alpha}{3}(M^2+2m_e^2)\sqrt{1-\frac{4m_e^2}{M^2}}\,.
\end{eqnarray}
The decay width $\Gamma(R\rightarrow N\gamma^*)$ is described within the extended VMD model~\cite{Kri01} in
terms of three transition form-factors (magnetic, electric and Coulomb) in case of a resonance with spin $J>1/2$ and 
two for $J=1/2$, which is just the number of independent helicity amplitudes for the respective spin value. 
The free parameters of the model are fixed by constraining the asymptotic form of the form-factors by quark counting rules~\cite{Bro73} 
and fitting to the experimental data for photo-production and electro-production amplitudes and partial-wave analysis for multichannel
$\pi N$ scattering.  The number of intermediate $\rho$ or $\omega$ states required to describe the transition form-factors depends
on the spin $J$ of the resonance in question: namely $J-1/2+3$. For the case that $J_{max}=7/2$ one needs 6 intermediate mesons,
with the masses chosen as follows: 0.769, 1.250, 1.450, 1.720, 2.150 and 2.350 (in GeV). Within this model a consistent description
of radiative and mesonic decays could be achieved. Further details about the extended VMD can be found in Ref.~\cite{Kri01}.

As already mentioned the decay widths $\Gamma(R\rightarrow N\gamma^*)$ are expressed in terms of the magnetic, electric and
Coulomb form-factors, more precisely they depend on the modulus squared of these form-factors. In the extended VMD each of these 
form-factors is in turn expressed  as a linear superposition of the contributions from the intermediate vector mesons ($\rho$ or $\omega$):
\begin{eqnarray}
G_{T}^{(\pm)}(M^2)=\sum_{k}\,\mathcal{M}_{Tk}^{(\pm)}
\end{eqnarray}
with $T$ standing for each of possible form-factors, $(\pm)$ 
denotes states of normal and abnormal parity respectively and 
the sum is over the intermediate mesons. The amplitude
\begin{eqnarray}
\mathcal{M}_{Tk}^{(\pm)}=h_{Tk}^{(\pm)}\,\frac{m_k^2}{m_k^2-im_k\Gamma_k-M^2}
\end{eqnarray}
represents the contribution of the $k^{th}$ vector meson to the 
amplitude of type $T$. The residues $h_{Tk}^{(\pm)}$ are fixed by the requirement
that the asymptotic expression of the form-factors is consistent 
with the quark counting rules~\cite{Bro73}. This leads to a destructive interference
between the intermediate vector mesons, since quark counting 
rules predict a behaviour steeper for the form-factors  than the $1/M^2$ contribution
of a single meson. In the  medium it is expected that the 
coherence between the contributions of individual mesons is at least partially lost. In the
extreme case of total decoherence, this would lead to the following 
replacement in the expression for the decay width,
\begin{eqnarray}
\textrm{\Large $\arrowvert$}\,\sum_k\: \mathcal{M}_{Tk}^{(\pm)}\,\textrm{\Large$\vert$}^2 
\longrightarrow \sum_k\:|\,\mathcal{M}_{Tk}^{(\pm)}\,|^2\,,
\end{eqnarray}
which will result in an enhancement of the resonance contributions. 
In reality both, the density
and wavelength of the meson are finite. Introducing the decay 
length $L_D$ of a resonance and its collision length $L_C$ one can determine
the probability of coherent decay ($i.e.$ the meson decay takes 
place before the first collision) as $w=\frac{L_C}{L_C+L_D}$. In order to
account for the decoherence effect, one can introduce an 
enhancement factor $E_T(M^2,\vec{Q}^2)$,
\begin{eqnarray}
\arrowvert\,G_T^{(\pm)}(M^2)\,\arrowvert^2\, \longrightarrow \,E_T^{(\pm)}(M^2,\vec{Q}^2)\,)\,
\arrowvert\,G_T^{(\pm)}(M^2)\,\arrowvert^2\,,
\end{eqnarray}
The dependence on the space-like part $\vec{Q}$ of the vector meson 
momentum originates from the definition of the collision length
$L_C$. Further details can be found in Ref.~\cite{She03}.

\subsection{The RQMD model}
The heavy-ion reaction dynamics is described within the 
framework of relativistic quantum molecular dynamics. 
The T\"ubingen (R)QMD transport code \cite{Uma98} has been 
extended to include all nuclear resonances with masses below 2 GeV, 
in total 11 $N^*$ and 10 $\Delta$ resonances.
A full list with the corresponding masses and decay widths to various 
channels can be found in Tables III and IV of Ref.~\cite{She03}. For the
description of dilepton production through baryonic resonances, 
respectively, the $\rho$ and $\omega$ production in $NN$ and $\pi N$
reactions, only the well established (4$^*$) resonances listed by 
PDG~\cite{PDG96} are taken into account. This corresponds to the same set
of resonances which has been used for the 
description of vector meson and dilepton production in elementary (p+p) 
reactions \cite{Fae03,Fuc03}, see also~\cite{She03,fuchs05}. As necessary 
for the present investigations, it provides e.g. an accurate 
reproduction of the measured pion yields in C+C reactions \cite{Stu01} 
within error bars. For the case of the $\eta$ meson, the 
fit of Ref.~\cite{Sib97} is used and therefore the production 
through the decay of nucleonic
resonances is completely neglected. A check of the two 
production mechanisms $NN\rightarrow NN\eta$ and $NN\rightarrow RN \rightarrow NN\eta$ 
has been performed leading to an almost similar $\eta$ yield in heavy ion reactions. 
$\eta$ absorption runs over the $N^*(1535)$ resonance. The corresponding 
 $\eta$ production cross sections in C+C collisions are consistent with 
the experimental results of Ref.~\cite{Ave03}.

\begin{figure*}
\centering
\includegraphics[width=16cm]{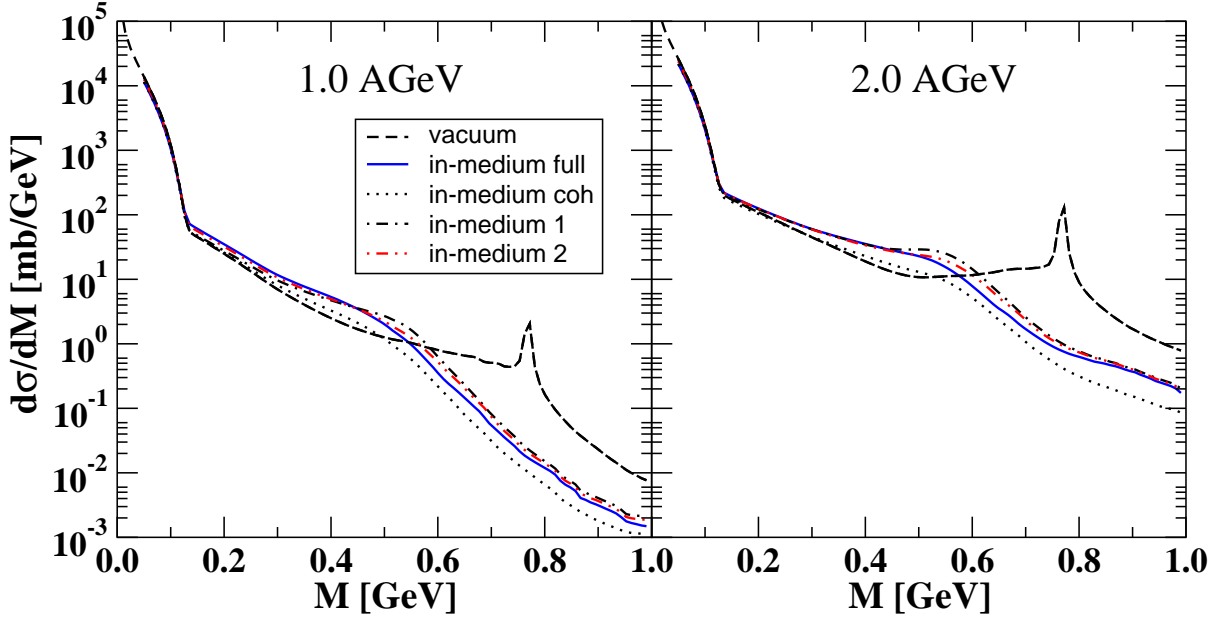}
\caption{\label{fig:dilcc1&2} Dilepton spectrum in 
C+C at 1 AGeV (left panel) and 2 AGeV (right panel). Besides the vacuum
calculation (dashed line) four different scenarios 
for in-medium modifications of the dilepton yield are presented. The
full in-medium calculation (full line) takes into account Brown-Rho 
scaling for the vector meson masses, collisional
broadening ($\Gamma^{\rm tot}_\rho$=250 MeV, $\Gamma^{\rm tot}_\omega$=125 MeV) 
and decoherence effects. The three other in-medium
calculations differ from the full one in the following respects: 
$\Gamma^{\rm tot}_\rho$=150 MeV (vacuum value) for the dashed-dotted curve, 
no decoherence effects for the dotted curve, and $\Gamma^{\rm tot}_\rho$=200 MeV together 
with $\Gamma^{\rm tot}_\omega$=60 MeV (double-dot-dashed curve).}
\end{figure*}

\section{Predictions for HADES}
It is expected that in nuclear matter the $\rho$ and $\omega$ 
mesons change their properties. Estimates for the collisional broadening
of $\rho$ in hadronic matter (nuclear matter or pionic gas) 
predict a collisional width of the order of the vacuum width. For the $\omega$
the vacuum width is only 8.4 MeV whereas in medium is 
expected to be more than one order of magnitude larger. The dilepton spectra at
intermediate energies, like those probed by the HADES 
and also the DLS experiments, are more sensitive to the $\omega $ meson collisional
broadening. In absence of such modifications the invariant mass dilepton spectrum 
would show a pronounced $\omega$ peak.  
In the DLS experiment such an enhancement 
has not been observed. Despite the limited 
mass resolution in~\cite{She03} an in-medium $\omega$ width of 
 $\Gamma_{\omega}^{tot}= 100\div 300$ MeV at nuclear matter density $\rho$ = 1.5$\rho_0$ 
has been extracted from 
the DLS data. The modification of the $\rho$ width was found to be similar in 
magnitude, i.e. $\Gamma_{\rho}^{\rm tot}=200\div 300$ MeV (again at $\rho$ = 1.5$\rho_0$). 
In the present calculations a linear density dependence of the $\rho$ and $\omega$ 
decay widths is 
assumed, i.e. $\Gamma_V^{\rm tot}=\Gamma_V^{\rm vac}+\rho/\rho_0 \Gamma_V^{\rm coll}$. 
 As an additional in-medium effect the masses of the mesons are supposed to vary 
as a function of the nuclear matter density at the point where 
the resonance decay occurs, following a Brown-Rho scaling law 
$m^{*}_{V}=m_V(1-\alpha \rho/\rho_0)$ with $\rho$ the local baryon density 
and $\alpha =0.2$. In contrast to \cite{She03} the 
dropping mass scenario is now included in addition to the 
collisional broadening and decoherence medium effects. 

A second medium effect is the due to decoherence, which mostly affects the dilepton
spectrum below the $\rho/\omega$ peak. The probability for coherent 
decay depends both on the in-vacuum decay widths (through $L_D$)
and the collisional broadening (through $L_C$). 
For the ground state vector mesons the following values for the
collisional widths have been adopted: $\Gamma_{\rho}^{\rm coll}$=100 MeV 
and $\Gamma_{\omega}^{\rm coll}$=116.6 MeV at $\rho=\rho_0$ with the
same values for their radially excited states that enter in 
the built up of the extended VMD. The vacuum widths of the radially excited
mesons are larger than those of the ground state $\rho$ and $\omega$ 
meson and as a consequence they tend to decay coherently. The
decoherence effect is largest for the $\omega$ vector meson 
since its vacuum width is very small.

The results for C+C collisions at 1 and 2 AGeV are shown in~\figref{dilcc1&2}.
Besides the dilepton spectrum with in-vacuum properties of the intermediate mesons 
(depicted by a dashed line) the effects of three different 
in-medium scenarios on the same spectrum are
also shown. The calculation in which all the in-medium effects 
are included to their full extent is
depicted by a full line. This case corresponds to maximal 
collisional broadening, i.e. $\Gamma_\omega^{\rm tot}$=125 MeV, $\Gamma_\rho^{\rm tot}$=250 MeV, 
both at   $\rho=\rho_0$, and it includes Brown-Rho scaling for the meson 
masses and decoherence effects.

The remaining three
calculations provide insight on the significance of the individual in-medium effects, even tough 
strictly speaking they cannot be disentangled. The variation of the $\rho$ meson 
width between $\Gamma_\rho^{\rm tot} =150 \div 250$ MeV 
leads to a modification of the dilepton yield by a 
factor of 2 in the dilepton mass range 0.5-0.8 GeV (compare the full and dashed-dotted curves). 
Decoherence effects in nuclear medium are 
responsible for at most a 50$\%$ change in the dilepton spectra at intermediate 
masses (dotted and full lines).

Some of the $\omega$ mesons, produced in the final stages of the collision or at the surface of
the interaction region might escape with vacuum properties and thefore lead to a small
peak in the dilepton cross-section. The density dependent meson widths include these 
possibilties. To explore the possibility of a reminiscent $\omega$ 
peak in more detail an additional calculation with a moderate in-medium $\omega$ 
width $\Gamma_\omega^{\rm tot}$= 60 MeV, together with $\Gamma_\rho^{\rm tot}$=200 MeV, 
(again at $\rho_0$) is shown. It should be noted that such a value for 
$\Gamma_\omega^{\rm tot}$ is in agreement with the analysis of \cite{cbelsa} from 
photo-nucleus reactions.  An increase of at most 50$\%$
in the dilepton yield is observed with respect to the full 'in-medium' 
scanario in the 0.4-0.8 GeV mass region, with no sign of a sharp peak. 

The effect of the Brown-Rho scaling is well known: it produces
a shift of the $\rho/\omega$ peak from its vacuum position towards 
lower dilepton invariant masses, namely around 0.6 GeV. The peak 
dissolves once the width of the $\omega$ meson is changed to 
its in-medium value. All results have been obtained with a strong  $N^*(1535)N\omega$ 
coupling as enforced by the fit of the resonance model 
parameter to nucleon resonance  electro- and photo-production \cite{Kri01} and which has been 
used in \cite{She03,Fuc03}. 

\begin{figure*}
\centering
\includegraphics[width=16cm]{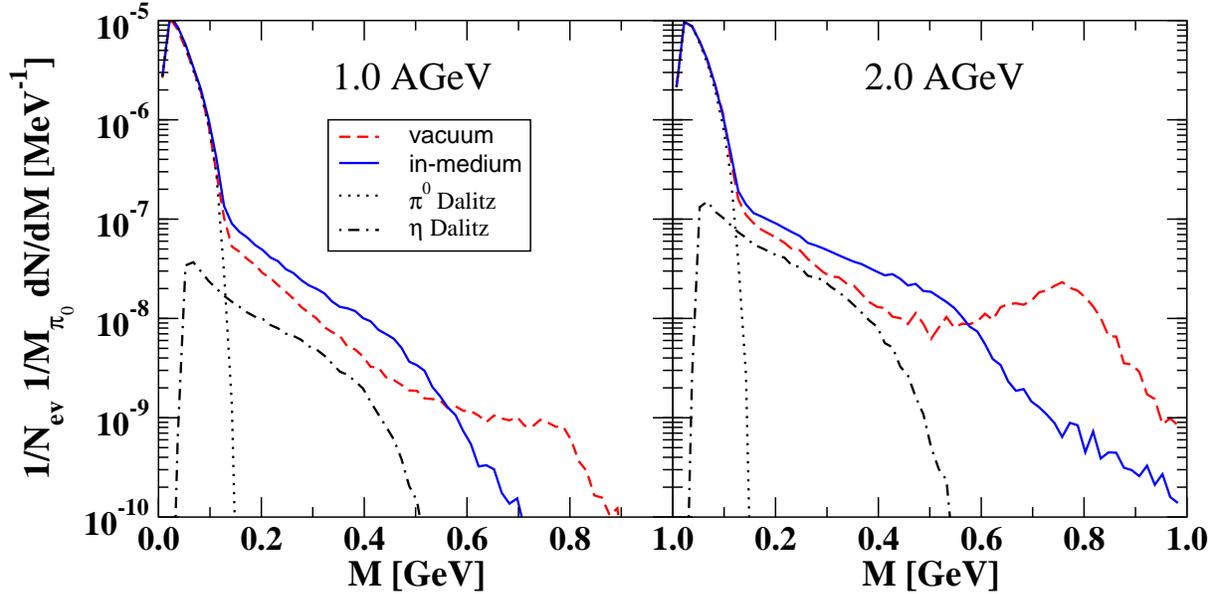}
\caption{\label{CCfilter_fig} Dilepton spectrum in C+C reactions at 1.0 and 2.0 AGeV 
after application of the full HADES acceptance filter. Calculations without 
(vacuum) and including in-medium effects (in-medium), i.e. maximal $\rho$ and 
$\omega$ collisional broadening, mass shifts and decoherence, 
are shown.}
\end{figure*}

The results of \figref{dilcc1&2} are pure theoretical results, $\it{i.e.}$ 
they have not been filtered in order to account for the experimental detector 
acceptance. Such a procedure is, however, indispensable for a meaningful 
comparison to data. In order to investigate up to what degree the HADES experiment will 
be able to discriminate between the different scenarios, we apply in the 
following the full HADES acceptance filter in combination with a 
smearing procedure for the corresponding HADES mass resolution. 
The filtered results are shown in in Fig. \ref{CCfilter_fig}.
The 'in-medium' calculation contains the combination of all 
medium effects under consideration, i.e.  $\rho$ and $\omega$ collisional broadening 
and mass shifts and decoherence (corresponding  to the full lines in~\figref{dilcc1&2}). 
The spectra are normalised to the number of events $N_{\rm ev}$ and to 
the $\pi^0$  multiplicity. Contributions from $\pi^0$ and $\eta$ Dalitz 
decay are shown separately. The difference between the 'vacuum' and  
the 'in-medium' calculation is still clearly visible: most pronounced 
are the medium effects in the mass region around the $\rho/\omega$ peak
($M\sim 0.6\div 1$ GeV) where a complete dissolution of the  $\rho$ and 
in particular the $\omega$ peak is predicted. This effect is even more 
pronounced at 2 AGeV and the HADES experiment will be able to clearly 
discriminate between the 'vacuum' and the 'in-medium' scenarios. 

\begin{figure}
\unitlength1cm
\begin{picture}(11.,8.0)
\put(0,0){\makebox{\epsfig{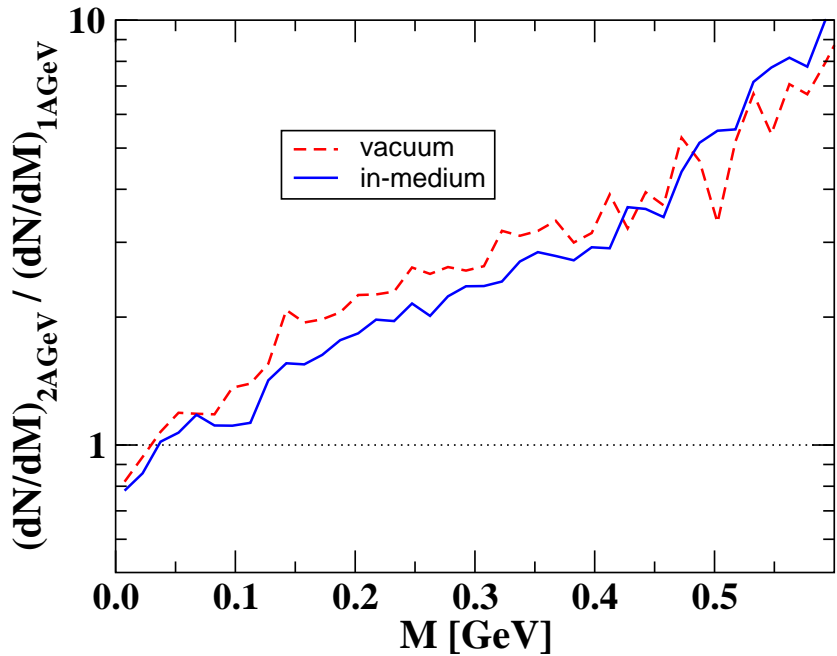}}}
\end{picture}
\caption{Ratio of the dilepton spectrum in C+C reactions at 2.0 over 1.0 AGeV 
after application of the full HADES acceptance filter. Calculations without 
(vacuum) and including in-medium effects (in-medium), i.e. $\rho$ and 
$\omega$ collisional broadening and mass shifts and decoherence, 
are shown.  
}
\label{fig:CCratio_fig}
\end{figure}

In the low and intermediate mass region the medium effects are less 
pronounced, i.e. of the order of about 50\%. 
Decoherence affects the dilepton pairs over almost the entire spectrum. 
It is, however,  the only source for 'in-medium' changes at low invariant dilepton masses,
below 0.4 GeV. To discriminate the various scenarios experimentally 
in the low and intermediate mass region will be difficult, at least 
in the light C+C system. However, keeping in mind that medium effects 
are often better visible at subthreshold energies as known e.g. 
from kaon production \cite{fuchs05}, it is natural to build 
the ratio between the spectra at 1 and 2 AGeV. This is done in 
~\figref{CCratio_fig} where the ratio of the dilepton spectra at  1 and 2 AGeV for the 
'in-vacuum' (dashed line) and 'in-medium' (full line) scenarios are  
plotted. Here aswell the theoretical predictions have been filtered using the HADES
acceptance filter which allows to infer the correctness of the conjecture 
concerning the relevance of such effects directly from experiment. 
From \figref{CCratio_fig} one observes a stronger in-medium enhancement 
of the low mass yield  at 1 AGeV compared to 2 AGeV which results in a 
smaller value for the ratio. The effect is, however at the 20-30\% level 
which requires very high 
resolution data. 

\section{Final conclusions}
In this paper we have presented predictions for the dilepton emission in 
heavy ion reactions at C+C at 1 and 2 AGeV. Experimental data for
these two reactions will be available in the near future from the 
HADES collaboration at GSI. A clear distinction between 
'vacuum' and 'in-medium' scenarios for $\rho$ and $\omega$ properties 
is possible in the mass region around the $\rho/\omega$ peak. In particular 
at 2 GeV the effect from an in-medium broadening of the vector mesons 
is dramatic and leads to a strong suppression of the spectrum. At low 
invariant masses the in-medium effects, in particular the  decoherence, 
are less pronounced, i.e. on the  20-30\% level, but can be expected to be 
more clearly seen in larger systems than C+C.  

We would like to acknowledge the help of our experimental colleagues from HADES who have 
filtered the results of our model with the HADES filter.


\end{document}